# Disorder control in crystalline GeSb$_2$Te$_4$ and its impact on characteristic length scales

Matthias Maximilian Dück[1], Tobias Schäfer[1], Stefan Jakobs[1], Carl-Friedrich Schön[1], Hannah Niehaus[1], Oana Cojocaru-Mirédin[1], Matthias Wuttig[1,2,†]

[1] 1. Physikalisches Institut (IA), RWTH Aachen University, 52056 Aachen, Germany
[2] JARA-FIT, RWTH Aachen University, Germany
[†wuttig@physik.rwth-aachen.de]

Crystalline GeSb$_2$Te$_4$ (GST) is remarkable material, as it allows to continuously tune the electrical resistance by orders of magnitude without involving a phase transition or stoichiometric changes, just by altering the short-range order. While well-ordered specimen are metallic, increasing amounts of disorder can eventually lead to an insulating state with vanishing conductivity in the 0K limit, but a similar number of charge carriers. These observations make disordered GST one of the most promising candidates for the realization of a true Anderson insulator. While so far the low-temperature properties have mostly been studied in films of small grain size, here a sputter-deposition process is employed that enables preparation of a large variety of these GST states including metallic and truly insulating ones. By growing films of GST on mica substrates, biaxially textured samples with huge grain sizes are obtained.
A series of these samples is employed for transport measurements, as their electron mean free path can be altered by a factor of 20. Yet, the mean free path always remains more than an order of magnitude smaller than the lateral grain size. This proves unequivocally that grain boundaries play a negligible role for electron scattering, while intra-grain scattering, presumably by disordered vacancies, dominates. Most importantly, these findings underline that the Anderson insulating state as well as the system's evolution towards metallic conductivity are indeed intrinsic properties of the material.

## I. INTRODUCTION

Phase change materials are characterized by a remarkable property portfolio. They can be rapidly and reversibly switched between an amorphous and a crystalline state [1,2]. This transformation is accompanied by an unconventional change of bonding mechanism [3,4,5] and a significant atomic rearrangement. The unusual combination of properties is attractive for various applications including rewriteable optical data storage, thermoelectrics and non-volatile electronic memories. For the latter two applications, a detailed understanding of heat and charge transport is very important.

Interesting enough, several crystalline phase change materials have been shown to possess highly unconventional transport properties for both charge and heat [6]. Depending on the order in the crystalline films, which can be tuned by a variation of deposition temperature or post-deposition annealing, several phase change materials show insulating behavior at low temperatures. In particular, compounds along the pseudo-binary line between GeTe and $Sb_2Te_3$ such as $Ge_2Sb_2Te_5$ or $Ge_1Sb_2Te_4$ (GST225 or GST124) can reveal such insulating properties. This is striking, since these materials possess a high density of electronic states at the Fermi energy. The resulting large carrier concentration should thus lead to metallic transport, i.e. a non-vanishing conductivity in the 0 K limit, contrary to experimental observation. Hence, charge transport is apparently controlled by exceptional levels of disorder, indicating that these materials might constitute a much sought-after state of matter, a true electronic Anderson insulator. Such a finding is quite remarkable, since in the last 50 years it has proven very challenging to identify crystalline solids where transport is exclusively governed by disorder, while electron – electron interactions (electron correlations) can be ignored. This difficulty has been explained on theoretical grounds, showing that increasing disorder also increases electronic correlations so that the realization of a true Anderson insulator has been deemed very daunting. To ensure that true disorder induced localization can be realized in certain crystalline phase change materials, it is hence highly desirable to carefully characterize all relevant length scales related to charge transport. This includes the elastic and inelastic electron mean free path and the length scales characterizing electron-electron and electron-phonon scattering as well as electron hopping. This is the main goal of the present study.

Previous studies have already explored the link between disorder in crystalline films of phase change materials and the resulting transport properties focusing on compounds along the pseudo-binary line between GeTe and $Sb_2Te_3$. Materials such as GST124 or GST225 are amorphous upon sputter deposition at room temperature. Annealing can bring them to a metastable (cubic, distorted rocksalt) and stable structure (trigonal phase, described by the rhombohedral space groups P-3m1 or R-3m) [7,8]. In both crystalline phases, the anion sublattice is populated by Te atoms, whereas Ge atoms and Sb atoms as well as vacancies share the cation sites. Depending on the composition, the amount of stoichiometric vacancies varies between 0% (GeTe) and 33% ($Sb_2Te_3$) of the cation sublattice [9]. If only such stoichiometric vacancies were present in the crystalline systems, these compounds would be narrow-gap semiconductors. Yet, they all possess large numbers of excess vacancies producing a high number of charge carriers (holes) at the Fermi level, which should lead to metallic transport [10].

The observation of a disorder-driven metal-to-insulator transition (MIT) upon increasing annealing temperature was first reported by [11] for crystalline GST124. They observed a sign change of the temperature coefficient of resistivity (TCR) upon increasing annealing temperature, which they attributed to a significant increase of order and hence electron delocalization upon annealing. Subsequent low temperature measurements [12] confirmed the existence of an MIT, yet also established that the TCR sign change and the MIT do not occur for the same annealing temperature, i.e. degree of crystalline disorder. Transport measurements revealed that the change in conductivity was not caused by a significant change in charge carrier density, but a very pronounced change of electron mobility instead. This already pointed to a prominent role of disorder. The number of charge carriers, on the contrary, was rather firmly linked to the number of vacancies [13], but hardly changed during annealing.

Zhang et al. [14] performed DFT calculations to identify the origin of charge carrier localization. They showed that the distribution of vacancies governs the transport properties. The prevalent structural rearrangement upon annealing is the accumulation of vacancies in vacancy layers (VLs) [15]. While randomly distributed vacancies were shown to produce localization centers for charge carriers, the VLs populated with vacancies became electrically inactive. Thus, vacancy ordering controls the electrical properties, and the pronounced mobility increase is due to the ordering on the cation sublattice.

The experiments discussed so far were performed on polycrystalline films with randomly oriented crystallites. Recently, the benefits of textured films to study the interplay between disorder and transport in GST materials have been demonstrated [16]. Features in the XRD diffraction pattern of GST225 were identified that are directly related to the formation of vacancy layers. The strength of this signal could be linked to the TCR, establishing a direct experimental link between vacancy ordering and the transport properties, which is highly desirable. However, the different samples investigated are obtained by very different processes involving amorphous deposition at room temperature as well as deposition on a heated substrate by MBE, and subsequent treatment involving post annealing and laser crystallization. Mio et al. [17] suggested based on their TEM studies that the fundamental differences in sample preparation may cause differences in the film's microstructure including the concentration of defects such as stacking faults. Additionally, changes of the overall microstructure, such as the grain size distribution of the films, seem likely. In a later study [18], an MBE-based annealing series of amorphously deposited material that created crystalline samples with pronounced texture was performed. Room temperature resistivity measurements showed a large spread in sample resistivity upon variation of the annealing temperature. These findings indicate that textured GST thin films provide unparalleled opportunities to unravel the interplay of structural perfection and transport properties. To realize the full potential of this unique material system, two different types of length scales need to be compared: the length scales that describe the film's microstructure, including the horizontal and vertical grain size; the length scales that describe the charge transport such as the elastic and inelastic electron mean free path, and the length scales characterizing electron hopping, electron-electron as well as electron-phonon scattering. This comparison is presented here.

## II. EXPERIMENTAL DETAILS

GeSb$_2$Te$_4$ films of approximately 10 nm thickness have been deposited by DC magnetron sputtering on freshly cleaved muscovite mica substrates. As a layered 2D solid, this substrate provides atomically smooth surfaces of macroscopic size, free of dangling bonds when cleaved along the (001) basal plane [19,20]. Prior to this study, the GST/mica system was thoroughly optimized employing a custom-built sputter tool [21]. This deposition system enables significant variations of gas pressure, target - sample distance and deposition rate, which were adjusted for optimal texture and smoothness of the GST films. Subsequently, a series of samples was produced, where only the substrate temperature was varied. This enabled the preparation of GST films with varying degrees of disorder.

The samples were then characterized using X-ray diffraction, employing symmetric $2\theta - \omega$ scans (T2T) as well as rocking curves (RC) of the main diffraction peaks. In order to determine the grain structure of the material, $\phi$-scans have been combined with the electron backscatter diffraction (EBSD) technique. As an additional measure of quality control, the smoothness of the films was confirmed by atomic force microscopy (AFM).

Hall-bar structures were produced from thin films by adding a protective capping material and gold contacts, followed by subsequent ion milling of the unprotected areas. Indium contacts were cold-pressed to attach wires. Subsequently, pulsed-DC- as well as AC-lock-in-measurements were employed to characterize the electronic transport properties of the films in a temperature range from 1.8 K to room temperature. R vs. T- and magnetoresistance-scans as well as Hall-effect measurements will be presented for different temperatures. Special care was taken to stay within the ohmic regime and to avoid spurious features of Joule heating.

For these transport measurements, several devices have been produced from films of the same sputter runs. Individual samples provide good temperature homogeneity on the surface as confirmed by the homogeneity of the film's resistivity across the sample. On the other hand, pre-studies employing irreversible thermochrome temperature measurement stickers revealed temperature differences of the order of 1K between the individual samples of one run, which lead to moderate variations of resistivity. The measured temperature differences stem from different mica substrate thicknesses after cleaving and lead to a denser set of samples to be analyzed. Nevertheless, all samples produced at a specific deposition temperature reveal the same qualitative characteristics.

## III. RESULTS AND DISCUSSION

### A. Structural properties

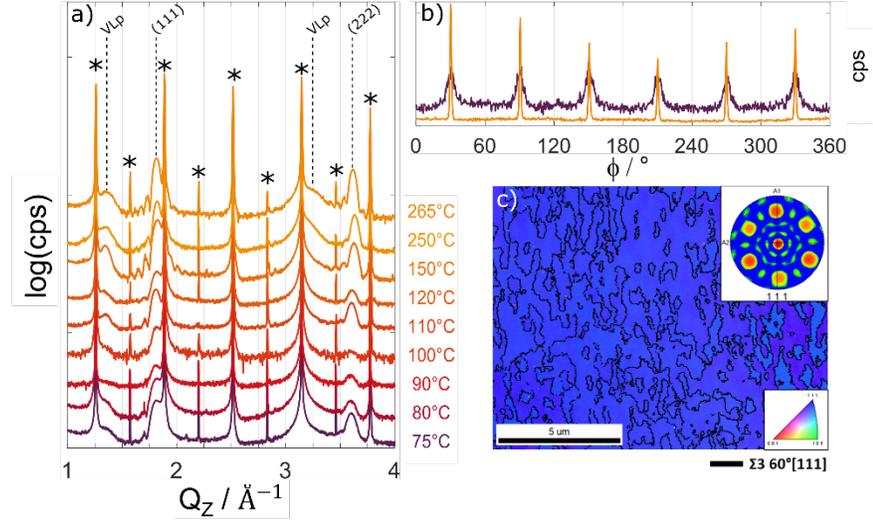

*Figure 1: (a) T2T-XRD scans of GST124 thin films at different deposition temperatures. Substrate peaks are marked by (\*). The peak family observed exclusively for this GST material is (00l), demonstrating that a pronounced film texture has been achieved. (b) ϕ-scans of the (107) reflex at the lowest and highest deposition temperature. Both scans show the six peaks expected for preferential in-plane orientation of crystallites with threefold rotational symmetry, which are subject to twinning. (c) EBSD inverse pole figure map of the 265°C sample. Only the two twinned domains separated by Σ3 twin boundaries with 60° [111] misorientation are observed.*

In order to compare electrical and structural properties of textured crystalline films, we start by characterizing their crystalline quality. T2T-scans of crystalline samples from the substrate temperature range between 75 and 265°C are presented in Fig. 1(a). The range was limited towards low temperatures by the crystallization threshold, while the material's desorption in vacuum at high temperatures set the high temperature limit. XRD scans in T2T geometry are sensitive to the thin film's lattice planes aligned with the substrate surface.

In Fig. 1(a), the different lattice planes are labelled. The numerous peaks marked by an asterisk (\*) belong to the mica substrate, which is probed along its long c-axis. For the peaks related to the GST films, we use the hexagonal notation for GST. This description simplifies the handling of XRD data in the case of the GST124 system, for which it is generally accepted to speak of a "cubic" and a "hexagonal" phase. The peaks around $Q_z$ = 1.8 and 3.6 Å$^{-1}$ ((0012) and (0024)) are significantly sharper and more intense than those at $Q_z$ ~ 1.4 and 3.3 Å$^{-1}$ ((009) and (0021)). Note that the former two peaks correspond to the (111) and (222) planes, if described in the cubic structure notation. The existence of additional, broader peaks with lower intensity (such as (009) and (0021)) has previously been observed in GST225 and named Vacancy Layer Peaks (VLP's) [16]. They were explained to originate from a partial population of the vacancy layers, which in the textured films produced here align parallel to the substrate. Therefore, the XRD diffraction patterns depicted can be ascribed to a single phase of cubic GST124 with partially ordered vacancy layers. In addition, Laue fringes are observed around the (0012) and (0024) diffraction peaks. They originate from coherent scattering of the x-rays from the lattice planes and can only occur for smooth films of high crystalline quality.

Subsequently, $\phi$-scans were performed to further characterize the film quality and grain structure as presented in Fig. 1(b) for the lowest and highest substrate temperatures. The samples of this series are clearly biaxially textured, showing the six peaks expected for a twinned threefold rotational symmetry. This conclusion is supported by an EBSD measurement of the sample deposited at 265°C shown in Fig. 1(c), where a large surface area larger than 10×10 µm$^2$ was mapped. The image clearly shows the biaxial texture including the twinned threefold rotational symmetry in agreement with the XRD data. EBSD detects the exclusive existence of ∑3 grain boundaries and determines the average grain size to about 1.5 µm. Such large grain sizes are ideal to minimize the impact of grain boundary scattering for the charge carriers, which is discussed in the second part of this paper.

From the XRD data discussed above, as well as additional XRR measurements, it is possible to derive several quantities directly related to the sample's microstructure. The corresponding numbers as well as their origin are presented in Tab. I.

The grain sizes are derived from the Scherrer equation, using the (0012) peak width of T2T scans (vertical) and rocking curves (lateral) and a Scherrer constant of K=0.9. This equation is based on the assumption that all broadening effects are ascribed to a limited grain size, and thus yield a lower limit estimate of the grain sizes. Therefore, we note that the numbers here are most probably underestimated, as broadening effects like substrate curvature and macroscopic step edges play a significant role for the layered, easily deformed mica substrate [19]. This is corroborated by the EBSD-based grain size determination of the 265°C sample, which yielded values in the range of µm.

The results displayed in Tab. I show an excellent agreement regarding data characterizing the grain size in out of plane direction. The vertical grain sizes deduced from the XRD peak width, as well as the coherence length from Laue fringe fits and the film thickness taken from XRR data determine similar values for each sample. Moreover, the lower limit estimate for the lateral grain sizes clearly exceeds the dimensions of grains in untextured samples, which were employed in earlier works [11]. Hence, we can conclude that all samples are monocrystalline in out of plane direction, and generally display an excellent crystalline quality, which does not seem to depend on the processing temperature.

| $T_{depo}$ (°C) | $d_{lat}$ (nm) | $d_{vert}$ (nm) | $d_{Laue}$ (nm) | $d_{XRR}$ (nm) |
|---|---|---|---|---|
| 75 | 340.2 | 7.4 | 8.3 | 8.0 |
| 80 | 217.5 | 8.3 | / | 8.0 |
| 90 | 217.5 | 8.2 | / | 8.2 |
| 100 | 427.1 | 7.9 | / | 8.3 |
| 110 | 117.8 | 9.0 | 8.7 | 8.8 |
| 120 | 166.3 | 8.5 | 8.0 | 8.4 |
| 150 | 78.5 | 9.8 | 9.0 | 8.9 |
| 250 | 327.2 | 12.1 | 11.4 | 10.8 |
| 265 | 376.5 | 12.2 | 11.1 | 10.0 |

*Table I: Structural length scales of GST124 samples deposited at various temperatures. $T_{depo}$ is the deposition temperature of the films, $d_{XRR}$ is the film thickness determined from the Kiessig fringes in XRR data, $d_{Laue}$ describes vertical coherence length obtained from Laue fringes of the (0012) XRD-T2T peaks, $d_{vertical}$ is a lower limit of the vertical grain size obtained by applying the Scherrer equation to the T2T FWHM of the (0012) peak, while $d_{lateral}$ is a lower limit of the lateral grain size obtained by applying the Scherrer equation to the rocking curve FWHM of the (0012) peak.*

B. Transport properties

With this detailed information of the film's microstructure, we can now explore the corresponding transport properties. They are depicted in Fig. 2 for films deposited at different temperatures. The room-temperature resistivity (Fig. 2(a)) can be tuned by more than two orders of magnitude, while the corresponding 2K conductivity even changes by more than five orders. Moreover, the samples produced at higher deposition temperatures depict clear metallic behavior in their R(T) curves, i.e. display a positive temperature coefficient of resistivity (TCR) and a non-vanishing conductivity in the 0 K limit. On the contrary, films deposited at lower temperatures show a negative TCR. Hence, samples produced at deposition temperatures of at least 90°C lead to metal-like TCR, while samples produced at 75°C and 80°C are insulator-like. The transition takes places for films having a room temperature resistivity of 2.7…3.8 mΩcm, consistent with previous findings [11].

While the TCR provides a first hint on transport properties, the behavior of a material at the lowest measurement temperatures is decisive. Scaling theory provides a range of methods and extrapolations to discriminate metals and insulators. They are typically designed to estimate if a finite conductivity would also remain in the 0 K limit. A widely accepted extrapolation [22,23,24] fits the conductivity with

$$\sigma(T) = \sigma_0 + \beta T^\nu, \quad (1)$$

where $\nu$ is usually assumed 0.5 and $\beta > 0$, while the constant's sign determines whether the material is metallic ($\sigma_0 > 0$) or insulating ($\sigma_0 < 0$).

For the samples analyzed here (see Fig. 2(b)), the lowest accessible temperature is about 2 K, since mica is a poor thermal conductor and the indium contacts become superconducting in this temperature range, and hence become bad heat conductors, too. From the temperature extrapolation towards 0 K, we conclude that all samples produced at 80°C are still metallic, whereas the samples produced at 75°C are true insulators. This result is corroborated by the method of logarithmic derivatives [25,26] (see SI Fig. S1). Hence, both extrapolations agree that the MIT occurs for room temperature conductivities between 71 and 103 S/cm, slightly higher than the value of 25-50 S/cm at which Volker et al. [12] observe the MIT.

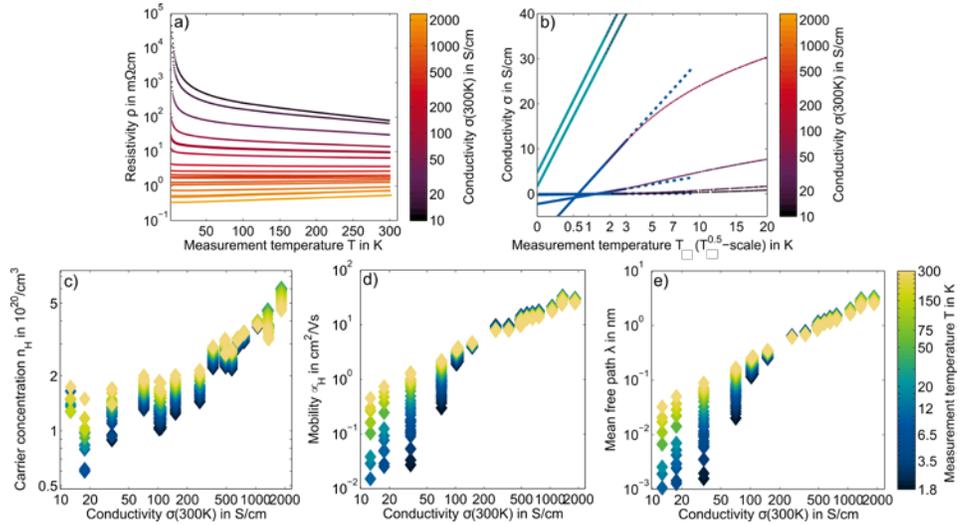

*Figure 2: Transport properties of the GST124-films produced at different temperatures. (a) Resistivity of the different samples as a function of temperature. Upon variation of the deposition temperature between 75 and 265°C, the room temperature resistivity changes by more than 2 orders of magnitude, while the resistivity at 2 K even changes by more than 5 orders of magnitude. (b) Low temperature extrapolation, revealing that only the samples produced at 75°C are real insulators, while the samples produced at slightly higher deposition temperatures have a negative TCR but a non-vanishing conductivity in the 0 K limit. (c) The concentration of charge carriers only increases by a factor of about 2.5, while the drastic increase of conductivity with annealing temperature can be*

*attributed to a similar increase in mobility of the charge carriers (Fig. 2(d)). (e) As a consequence, the mean free path increases by several orders of magnitude as well.*

Hence, similarly to [11,12] we can categorize our samples into four groups: while deposition temperatures around 70°C still lead to amorphous samples that are of no interest in the present study, deposition temperatures around 75°C lead to well textured but electrically insulating samples. A temperature window around 80°C allows for samples that are metals as defined by finite 0 K conduction, but behave as insulators in terms of their TCR (negative TCR). Finally, deposition between 90°C and 265°C produces (bad) metals with varying degree of disorder but clear metallic signature.

Combining the conductivity data with the Hall-effect allows for a deeper understanding of the change in transport properties. While the number of Hall carriers (Fig. 2(c)) only increases by a factor of about 2.5 with increasing deposition temperature, the Hall-mobility increases by more than two orders of magnitude as evaluated at room temperature. This is typical for disorder-induced localization, yet data for the Hall carrier concentration and mobility for highly insulating materials have to be interpreted with care. The data depicted in Fig. 2 show a number of interesting trends. While samples with conductivities below 300 S/cm display a Hall mobility which increases with measurement temperature (Fig. 2(d)), samples with higher conductivity reveal a mobility which decreases with increasing temperature. The former behavior is characteristic for temperature-activated transport, while phonon scattering dominates for the more conductive samples. Interesting enough, the TCR sign change coincides with the change in the main scattering mechanism. Thus, the TCR sign change is strictly speaking not a quantification of the MIT, but identifies the point where phonon scattering becomes more relevant for the material than temperature activation, which defines transport via localized states.

A temperature activation effect can also be seen for the concentration of charge carriers. This is reasonable, as the Hall-effect determines the number of mobile charge carriers. Notably, this temperature activated fingerprint can also be seen in metallic samples and persists even into the region where the TCR is metal-like. This emphasizes again that the TCR sign change is not a fundamental transition but describes the point where one effect (phonon scattering) starts to outperform a second one (temperature activation of charge carriers). Only the most conductive samples do not show an increase of charge carriers with temperature but show a linear decrease instead (see SI Fig S2).

From Hall- and conductivity measurements we can determine the mean free path for the charge carriers. This scattering length takes into account all scattering events independently of their type. Furthermore, the low and high conductivity samples adopt different trends in their temperature behavior (again with the TCR sign change as a demarcation line). While the low-mobility samples show a strong increase in mean free path with temperature, which is again a typical sign for temperature activation, the high mobility samples exhibit a marginal decrease in mobility. This slight decrease shows that for high measurement temperatures it is not only the defects but also a small share of phonon scattering that limits the mean free path. Finally, it should be noted that the least conductive samples imply scattering lengths below the size of an atom at low temperatures. This is indicative of the breakdown of the model of free carrier transport. Instead, under these conditions transport takes place via hopping, as will be elaborated later.

The measurements that have been presented so far can only reveal the scattering length of the most prominent scattering channel, but typically electrical transport in a material is governed by the interplay of several scattering lengths. This is why further measurements and evaluations have been performed, namely magnetoresistance for the metallic samples and hopping length evaluations for the insulating specimen.

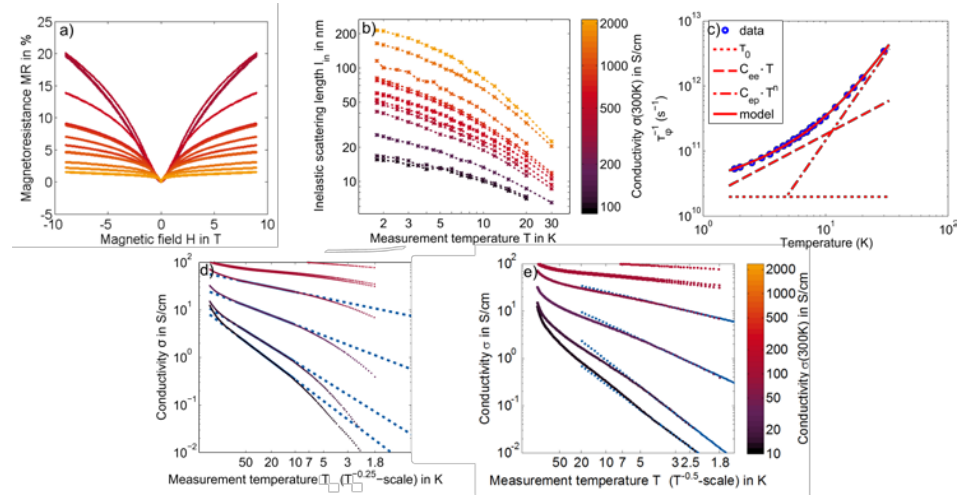

*Figure 3: (a) The magneto-resistance curves of the metallic samples depict weak anti-localization, which is strongest for the least metallic samples. (b) Fitting such curves is possible according to HLN-theory and yields inelastic scattering lengths. (c) The temperature dependency of the corresponding scattering rates $\tau_\phi^{-1}$ (as given by several magneto-resistance curves at different temperatures on the very same sample) allows for additional differentiation of the contributions of electron-electron ($\propto T$) and electron-phonon scattering ($\propto T^n, n = 2 \ldots 4$). (d,e) By contrast, the decisive length scale in insulating samples is the localisation length for hopping processes that are Mott-variable range hopping for relatively low temperatures and Efros-Shklovskii-hopping for even lower temperatures.*

As depicted in Fig. 3(a), the magneto-resistance of the metallic samples can be analyzed to get further insight into the metallic samples. As expected from previous studies [27] the magneto-resistance can be attributed to weak anti-localization. Weak anti-localization is one of the localization effects that are typical for disordered materials where pronounced spin-orbit coupling is present, in this case due to heavy atoms. High amounts of defect scattering allow some of the atoms to travel on closed loops, where self-interference of the two time-reversed paths leads to increased back-scattering, that is called weak localization. This effect is reversed by spin-orbit coupling that modifies the periodicity of the involved wave-functions and causes anti-localization. A magnetic field can suppress both localization and anti-localization. Although here the atomic composition remains unchanged, the magnitude of the anti-localization effect changes tremendously. This is caused by a reduction of defect scattering for higher temperatures, so that the magneto-resistance can be used as a probe for vacancy ordering as well.

Fitting the magneto-resistance according to a theory proposed by Hikami, Larkin and Nagaoka (HLN) [28] allows to quantify additional scattering lengths. Strictly speaking, the theory allows fitting only in case of the disorder parameter $k_F \lambda > 1$ (compare SI Fig S3), where $k_F$ is the Fermi wave vector, while $\lambda$ is the mean free path. Here, fitting yields decent representations of the data also for $k_F \lambda \lesssim 1$ (the three least conductive metallic samples), so that we will show those results as well. In this study, the inelastic scattering length is of highest interest (see Fig. 3(b)). The inelastic scattering length decreases with temperature as the underlying scattering processes are expected to become more prominent for increasing temperature. Furthermore, all temperature dependencies have the same qualitative shape, which underlines that also the underlying mechanisms are the same for all samples, although their TCR is different.

Interesting enough, the absolute values of the inelastic scattering depend on deposition temperature. To observe this more clearly, one can divide the inelastic scattering into electron-electron (ee) and electron-phonon (ep) contributions (see Fig. 3(c)). Obviously, we need to exchange the picture of length scales for scattering rates to be able to use Matthiessen's rule. In this case, we assume

$$\tau_\phi^{-1} = \tau_0^{-1} + \tau_{ee}^{-1} + \tau_{ep}^{-1} = \tau_0^{-1} + C_{ee}T + C_{ep}T^n; n = 2 \ldots 4, \qquad (2)$$

where the reason for a (small) temperature independent $\tau_0^{-1}$ contribution to scattering is still under debate [27]. The results of these fittings (see Fig. 4) show, that electron-phonon scattering is of similar magnitude in all samples, while electron-electron scattering decreases for more ordered samples. This appears counterintuitive at first sight, since the less ordered samples also show fewer carriers, but apparently the number of carriers is not the decisive factor here. By contrast, our findings can be summarized that more localized electrons lead to increased electron-electron scattering.

For the insulating samples the mean free path is not a meaningful characteristic for transport. Instead it is possible to obtain a localization length from hopping (compare [12] for the relevant models). The localization length explains how large the extent of the localized wave functions is. Fig. 3(d,e) shows the relevant transport data and fitting. The insulating samples are consistent with Mott-variable range hopping for relatively low temperatures and Efros-Shklovskii-hopping for even lower temperatures. In consequence, the relevant hopping length at 2 K is the Efros-Shklovskii one: It increases from 4.7 and 5.1 nm for the most insulating samples towards 13.3 and 51.8 nm, which even exceeds the film thickness. This steady increase in hopping length as well as the enormously high values are not surprising, as the samples are approaching the MIT.

## III. LENGTH SCALES OF THE GST SYSTEM

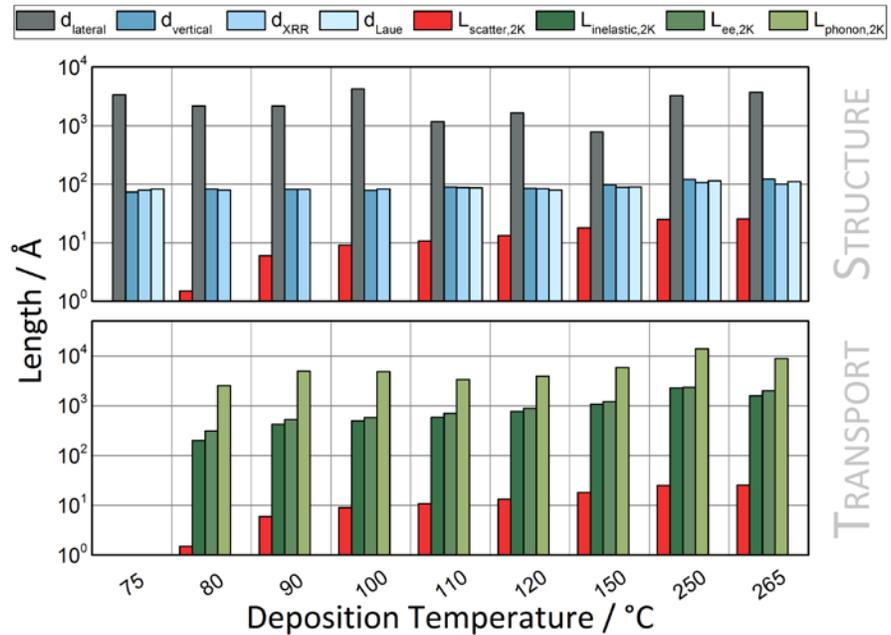

Figure 4: Bar chart comparing the length scales of structural (top) and transport measurements (bottom). The elastic scattering lengths is plotted in both charts, as this measure is directly comparable to structural lengths. It should be noted that no transport length scales are displayed for the 75°C sample, which as a true insulator cannot be meaningfully described within the same theoretical framework as the metallic samples.

Both the results from structure and transport measurements, which have been presented and discussed above, may readily be expressed in terms of length scales. In order to grasp the fundamental connection between the various length scales of the GST system, we mold the acquired information of structure and transport into a single picture.

The comparison of structure and transport properties is presented in Fig. 4. As stated before, all samples from this deposition temperature series are biaxially textured, which renders the number of random angle grain boundaries very low.

The lateral grain sizes, even in the lower limit estimate explained above, clearly exceed a size of 100 nm, and do not show a clear trend with deposition temperature. Within error bars, the vertical grain sizes are in good agreement with the vertical coherence length derived from fits of the Laue fringes, as well as the film thickness as measured by means of XRR. This demonstrates that the GST films are monocrystalline in out-of-plane direction, since the grains are shown to be as large as the film. Hence, the sample series studied here shows extraordinary crystalline quality, with biaxial texture, out-of-plane mono-crystallinity and lateral grains of several 100nm, independent of deposition temperature.

This enables us to ascribe all significant changes in transport properties to conditions that are intrinsic to the material, and not a property of the thin film, i.e. vertical and lateral grain size. Indeed, the elastic scattering length at 2 K, which can be ascribed to impurity scattering, is considerably smaller than the film thickness for all samples, rendering the material three dimensional in terms of transport. Even more importantly, the elastic scattering length is orders of magnitude smaller than the lateral (and also vertical) grain size, rendering defect scattering a clear intra-grain effect. The increase of the elastic scattering length with deposition temperature by more than an order of magnitude adds to this picture of an intra-grain effect, as structural effects such as changes in grain size and grain boundary angles can be excluded with the help of X-ray diffraction and EBSD. It is self-evident to ascribe these changes to point-like differences in the local atomic surroundings. Zhang et al. confirmed that only the degree of vacancy ordering in the material can have significant impact on mobility and localization, while the ordering of the Ge/Sb atoms on the cation sublattice is of minor importance. Thus, we conclude that we mainly change the degree of intrinsic vacancy order within this samples series.

The inelastic transport length scales should not be directly compared to structural length scales, as they describe a diffusive motion rather than a ballistic travel of the electrons. Nevertheless, a qualitative comparison is insightful as well: in the samples analyzed here the electron-phonon scattering length is neither dependent on the elastic scattering length nor on crystalline quality, which may be inferred from the lateral grain size. Instead, the differences in electron-phonon scattering length seem to be a statistical variation and originate from the precision of the determination of these least prominent scattering events. Electron-electron scattering is the more frequent inelastic scattering event at 2 K and hence its scattering length has been determined with higher precision. Electron-electron scattering clearly becomes more frequent for the more disordered samples that are produced at lower temperature. This might be linked to the disorder induced electron-electron interaction being suppressed for more ordered samples [29], as electrons are forced to stay close to each other by frequent elastic backscattering for the more disordered samples.

In summary, it has been demonstrated how one can identify, isolate and tailor the important length scales of the GST system, by producing samples that are highly ordered in terms of their X-ray patterns and X-ray coherence length, but can be insulating in terms of their electronic transport properties and electron mean free path.

**ACKNOWLEDGEMENTS**

The authors thank Marc Pohlmann for performing the EBSD measurements. This study has been supported by Deutsche Forschungsgemeinschaft (DFG) in the framework of SFB 917 "Nanoswitches". The research leading to the experimental results has received funding from the European Union Seventh Framework Programme (FP7/2007-2013) under grant agreement no. 340698.

**SUPPLEMENTARY INFORMATION**

**Figure S1: Logarithmic derivatives.** As described in the paper, the method of logarithmic derivatives is an alternative way to distinguish metals and insulators. The results obtained by this method are consistent with the conclusions obtained in this paper. Nevertheless, the analysis employing the method of logarithmic derivatives would benefit from data at lower measurement temperature. Although desirable, this lower temperature data is not feasible due to restrictions imposed by the choice of substrate and sample production method, as neither the mica substrate nor the cold-pressed indium contacts (that become superconducting) are viable thermal conductors at low cryogenic temperatures.

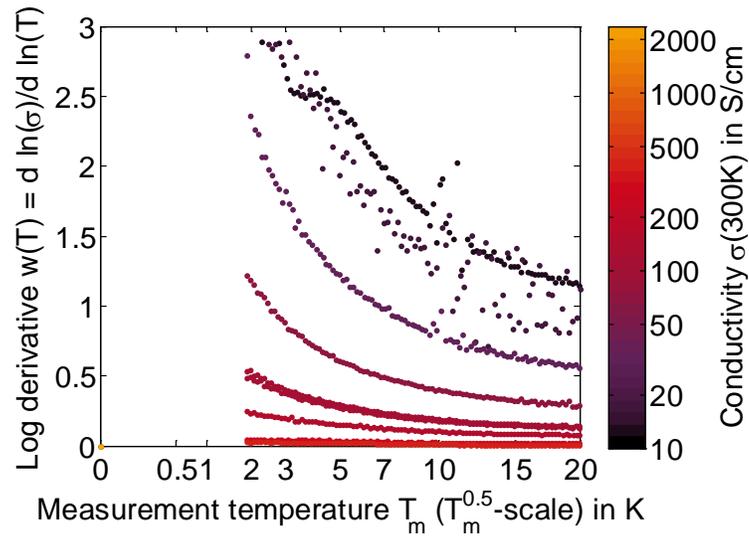

**Figure S2: Hall-effect vs. temperature.** In contrast to figure 2b) that illustrates the evolution of the Hall carrier concentration with measurement temperature in a rather reticent way, this plot explicitly displays this trend. A physical interpretation of the effects visible can be found in the discussion of fig 2b).

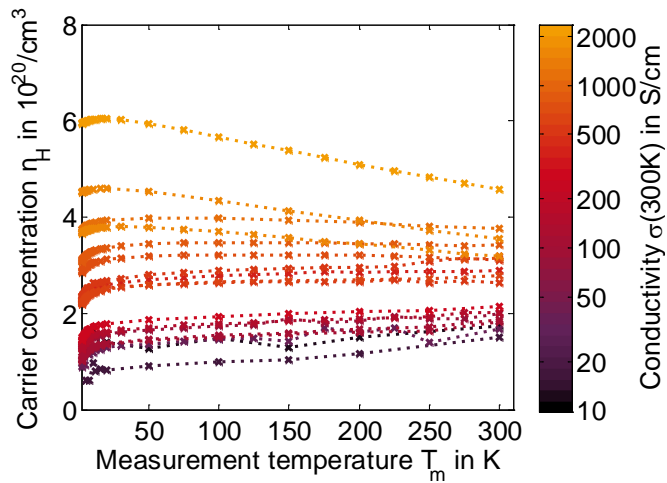

**Figure S3: Disorder parameter.** An additional parameter that can be calculated from classical transport theory is the disorder parameter that combines the mean free path with the Fermi wave vector.

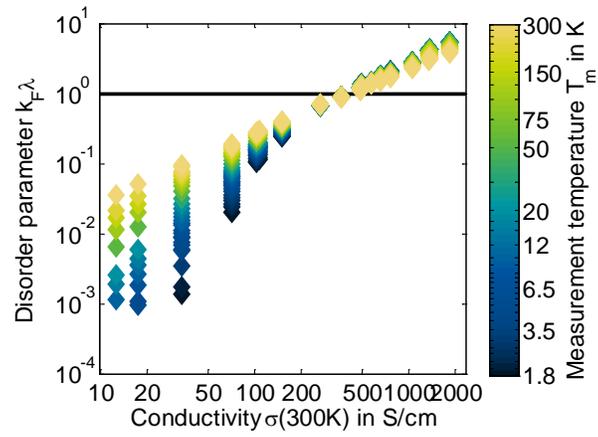